\documentclass[twocolumn,preprintnumbers,amsmath,amssymb,prb,superscriptaddress]{revtex4-2}

\usepackage{epsf}
\usepackage{graphicx}
\usepackage{sidecap}
\usepackage{soul}
\usepackage{array}
\usepackage{amsmath}
\usepackage{amssymb}
\usepackage{dcolumn}
\usepackage{epstopdf}
\usepackage{bm}

\usepackage{gensymb}
\usepackage{color}

\sethlcolor{green}
\usepackage{bbding}
\usepackage[T1]{fontenc}
\usepackage[utf8]{inputenc}

\usepackage[urlcolor=blue,colorlinks=true,citecolor=blue,linkcolor=blue,pdfstartview={FitH},bookmarks=false]{hyperref}

\usepackage[dvipsnames]{xcolor}
\definecolor{mygreen}{rgb}{0.0, 0.6, 0.0}
\definecolor{pjorange}{rgb}{0.8, 0.3, 0.0}
\definecolor{jlblue}{rgb}{0.2, 0.5, 0.7}

\def \beq {\begin{equation}}
\def \eeq {\end{equation}}
\def\bibsection{\refname}
\renewcommand{\refname}{\noindent\textbf{References}}

\begin{document}

\title{Observation of the Optical Phonons in  $\alpha$-MnTe films}

\author{Himanshu Sheokand}
\affiliation{Department of Physics, University of Central Florida, Orlando, Florida 32816, USA}

\author{Arun K Kumay}
\affiliation{Department of Physics, University of Central Florida, Orlando, Florida 32816, USA}

\author{Mazharul Islam Mondal}
\affiliation{Department of Physics, University of Central Florida, Orlando, Florida 32816, USA}

\author{Milo Sprague}
\affiliation{Department of Physics, University of Central Florida, Orlando, Florida 32816, USA}

\author{Ravinder Sharma}
\affiliation{Department of Physics, University of Central Florida, Orlando, Florida 32816, USA}
\affiliation{Nanoscience Technology Center, University of Central Florida, Orlando, Florida 32826, USA}

\author{Jayan Thomas}
\affiliation{Department of Physics, University of Central Florida, Orlando, Florida 32816, USA}
\affiliation{Nanoscience Technology Center, University of Central Florida, Orlando, Florida 32826, USA}
\affiliation{Department of Materials Science and Engineering, University of Central Florida, Orlando, Florida 32816, USA}

\author{Dariusz Kaczorowski}
\affiliation{Institute of Low Temperature and Structure Research,
Polish Academy of Sciences, Ok\'olna 2, 50--422 Wroc{\l}aw, Poland}

\author{Andrzej Ptok}
\email{aptok@mmj.pl}
\affiliation{Institute of Nuclear Physics, Polish Academy of Sciences}

\author{Madhab Neupane} 
\email{madhab.neupane@ucf.edu}
\affiliation{Department of Physics, University of Central Florida, Orlando, Florida 32816, USA}

\date{\today}

\begin{abstract}
The altermagnetic materials have emerged as model systems for studying spin-split electronic structures, yet controlled epitaxial growth on technologically relevant substrates remains challenging. Among the known candidates, MnTe stands out as a prominent altermagnetic material owing to its layered structure and high N\'eel temperature. Here, we report the molecular beam epitaxy (MBE) growth of high-quality $\alpha$-MnTe thin films on GaAs(111)B substrates and provide a comprehensive analysis of the growth evolution and structural properties. Raman spectroscopy reveals multiple vibrational features of $\alpha$-MnTe including modes near 121, and 140~cm$^{-1}$. Combined with first-principles phonon calculations, these features are identified as the Raman-active phonons of the hexagonal NiAs-type lattice. Our results show that the high crystalline quality of MBE grown $\alpha$-MnTe enables the complete experimental resolution of all symmetry-allowed Raman-active phonon modes, highlighting epitaxial $\alpha$-MnTe as a robust thin-film platform for investigating altermagnetism and its lattice-coupled excitations. 
\end{abstract}

\maketitle

Magnetic materials are central to modern spintronics, where their electronic and symmetry properties govern functionalities ranging from information storage to quantum technologies~\cite{Furdyna1988,Zutic2004,Hirohata2020,Baltz2018}. 
Traditionally, research has focused on ferromagnets~\cite{Wolf2001,Parkin2008} for their spin-polarized transport and anomalous Hall effect~\cite{Nagaosa2010,Sinova2015} and on antiferromagnets~\cite{Jungwirth2016,Baltz2018} for their compensated spin order with potential for fast and robust spin dynamics~\cite{Jungwirth2016,Baltz2018,Gomonay2014}. 
Recently, a new class of material known as altermagnets has been proposed and realized~\cite{aa,aaa,aaaa,Krempasky2024, b}. 
These systems break combined crystal and spin symmetries in such a way that electronic bands exhibit momentum-dependent spin splitting without requiring a finite magnetization~\cite{b,bb,bbb}. 
As a result, altermagnets~\cite{Guo2023,Mazin2023,Regmi2024,Dong2025,Sakhya2025,Sprague2025} combine the advantages of both antiferromagnets and ferromagnets.  

Among the known candidates, hexagonal $\alpha$-MnTe has emerged as a prototypical altermagnetic material. It crystallizes in the NiAs~\cite{Janik1995,ElectricalProperties2016} structure with collinear antiferromagnetic order and a N\'eel~\cite{Osumi2024,Roessler2025} temperature ($\sim310$ K) close to room temperature. Its electronic structure features alternating spin splitting~\cite{Gonzalez2024,Osumi2024band} along high-symmetry directions, leading to unconventional transport responses such as the anomalous Hall effect and anisotropic magnetoresistance\cite{d,e, cc}. Recent experiments have revealed that these responses are strongly sensitive to lattice strain, substrate choice, and growth parameters. For example, $\alpha$-MnTe films grown on lattice-matched substrates ~\cite{Kommichau1986,Bangar2025, Chen2026} behave differently compared to those on lattice-mismatched GaAs(111)~\cite{c,cc,eee}.  

Despite this rapid progress, important questions remain regarding the microscopic mechanisms that connect epitaxial growth, crystal symmetry, and altermagnetic phenomena. In particular, clarifying how homogeneous, phase-pure $\alpha$-MnTe thin films can be stabilized on GaAs(111)B by MBE remains an important unresolved issue. At the same time, the optical phonon response of epitaxial $\alpha$-MnTe, and how it differs from bulk $\alpha$-MnTe, remains largely unexplored.

\begin{figure*} 
	 
    \includegraphics[width=\linewidth]{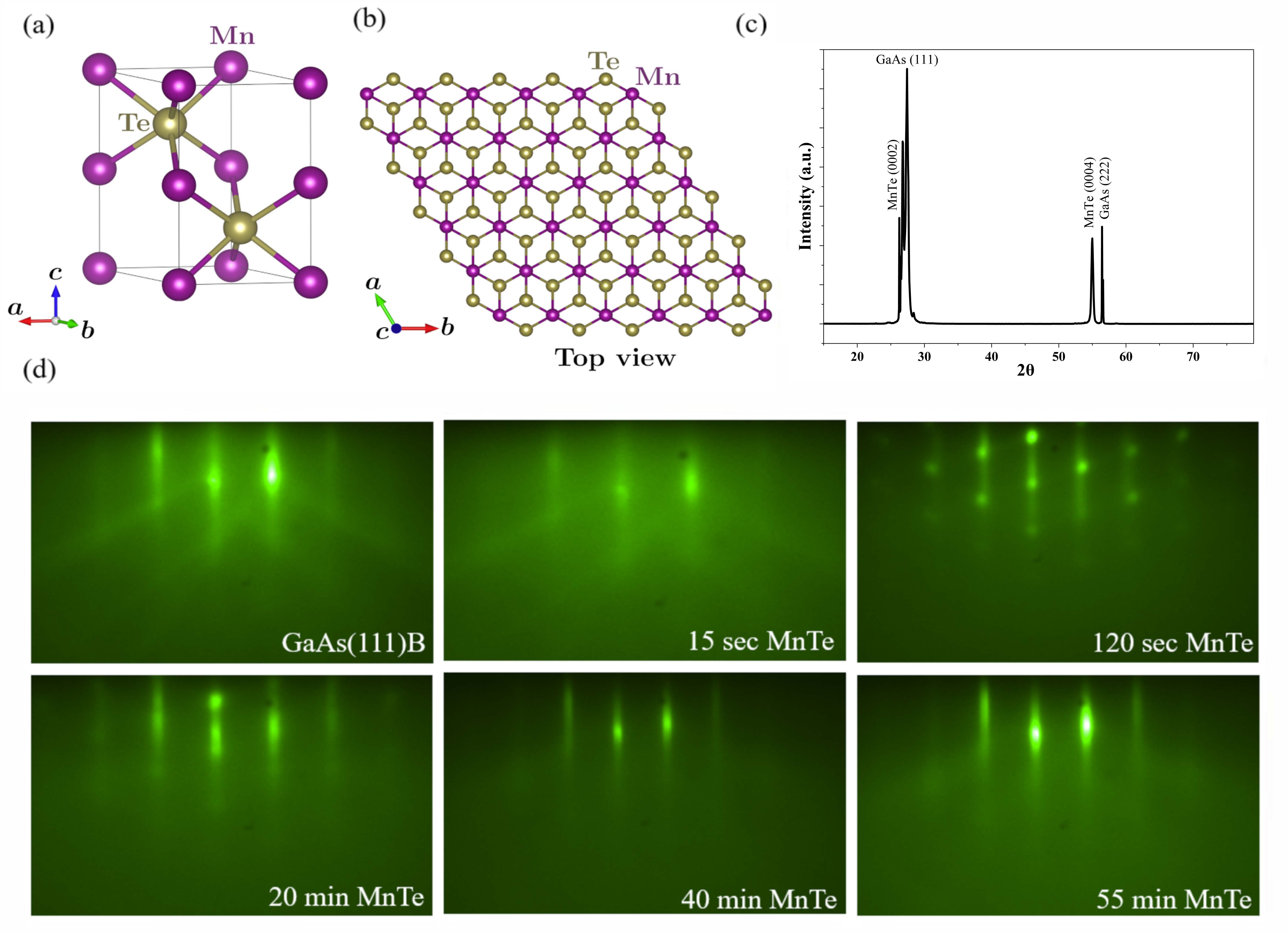}
	\caption{(a,b) Crystal structure of hexagonal $\alpha$-MnTe shown along two different crystallographic directions, illustrating the atomic arrangement and stacking sequence of alternating Mn and Te layers consistent with the NiAs-type structure. (c) X-ray diffraction pattern plotted as intensity versus $2\theta$ scan over the range of 10°–80°, showing the GaAs(111) and GaAs(222) substrate reflections together with the MnTe (0002) and (0004) film peaks. The observation of only \ensuremath{(000\ell)} MnTe reflections confirms highly oriented growth with the c-axis of $\alpha$-MnTe aligned normal to the GaAs(111) substrate surface. (d) RHEED patterns recorded during the growth of epitaxial $\alpha$-MnTe on GaAs(111)B, illustrating the evolution of surface structure and crystalline ordering throughout the deposition process.
    }
\label{fig.fig1}
\end{figure*}

In this work, we present an analysis of the synthesis of $\alpha$-MnTe thin films on GaAs(111)B using molecular beam epitaxy (MBE). We chose GaAs(111)B substrates for the growth of $\alpha$-MnTe films to intentionally disrupt the pristine crystal structure despite the lattice mismatch, as growth on lattice-matched substrates is possible but does not yield the same exotic electronic properties. However, only a few studies on the MBE growth of $\alpha$-MnTe are available in the previous studies~\cite{c,cc,eee, Bey2025} because of its complex growth mechanism and precise parameter control. Here, we emphasize the role of substrate temperature in achieving uniform, well-oriented layers of $\alpha$-MnTe films on GaAs(111)B substrates via MBE. The growth rate was calibrated with a quartz crystal microbalance, and the entire growth process was monitored \textit{in situ} by reflection high-energy electron diffraction (RHEED). By systematically optimizing the Mn:Te flux ratio and varying the substrate temperature, we achieved high-quality epitaxial $\alpha$-MnTe thin films with thicknesses of $\sim$ 40 nm. Growth was carried out in the temperature of 425$^{\circ}$C, where uniform, single-phase films with smooth topography and stoichiometric composition were obtained. Highly oriented and continuous films were realized despite the substantial lattice mismatch. A detailed ex-situ study of the synthesis was performed using X-ray diffraction (XRD), scanning electron microscopy (SEM), energy-dispersive X-ray spectroscopy (EDX) were employed to characterize the structural phase, surface morphology, and elemental composition of the as-grown films. Our findings from detailed Raman analysis resolves the characteristic phonon modes of hexagonal $\alpha$-MnTe, unambiguously establishing the NiAs-type phase and the high crystalline quality of the epitaxial films. At the same time, it reveals that the vibrational response of epitaxial $\alpha$-MnTe thin films is fundamentally distinct from that of bulk $\alpha$-MnTe. Details of the thin-film growth, computational methods, X-ray diffraction, scanning electron microscopy, and Raman spectroscopy measurements are provided in Sec. 1 of the Supplemental Material (SM).

\begin{figure*} 
	\includegraphics[width=\linewidth]{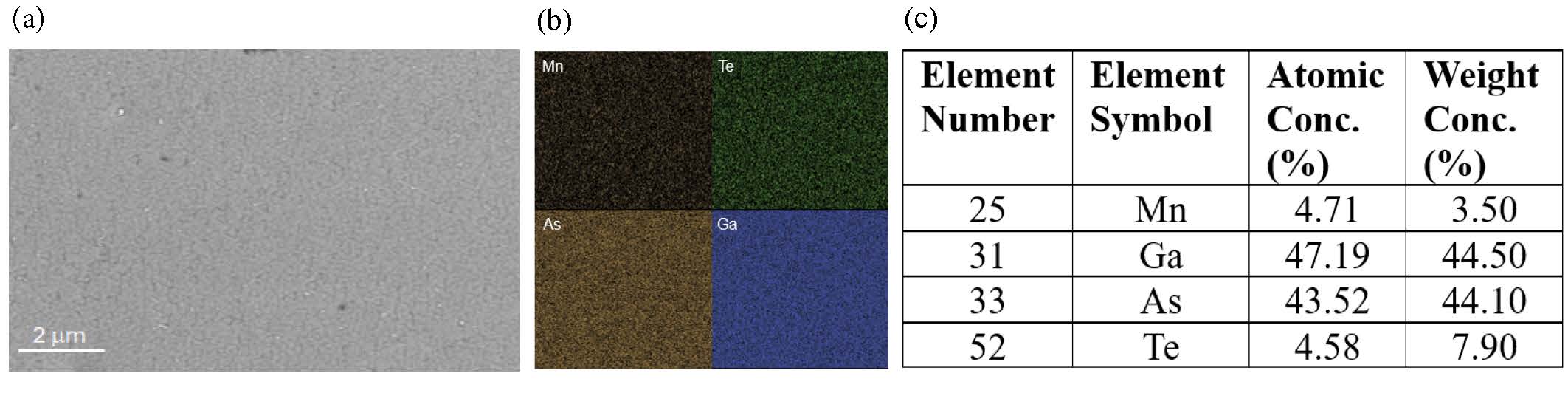}  
	\caption{{Scanning electron microscopy (SEM) and energy-dispersive X-ray spectroscopy (EDX) analysis of $\alpha$-MnTe/GaAs(111)B. (a) SEM image of a selected region of the 10~$\mu$m MnTe film where the EDX spectra were acquired. (b) Elemental maps show uniform Mn and Te distribution along with strong Ga and As signals from the substrate. (c) Quantitative EDX yields a near-ideal stoichiometric Mn:Te ratio ($\sim$1:1).
}} \textbf{} 
\label{fig.fig2}
\end{figure*}

Fig.~\ref{fig.fig1} summarizes the structural characterization of the epitaxial $\alpha$-MnTe thin film grown on GaAs(111).
Figs.~\ref{fig.fig1}(a) and~\ref{fig.fig1}(b) shows the crystal structure of hexagonal $\alpha$-MnTe, illustrating the atomic arrangement and stacking sequence viewed along two different crystallographic directions. The structure consists of alternating Mn and Te layers arranged in a hexagonal lattice, consistent with the NiAs-type structure. Fig.~\ref{fig.fig1}(c) presents the XRD pattern plotted as intensity versus 2$\theta$ scan collected over the range 10°–80°, confirming the high crystallinity and phase purity of the film. The diffraction pattern exhibits pronounced substrate reflections corresponding to GaAs(111) and GaAs(222), along with clear MnTe (0002) and (0004) reflections. The exclusive appearance of (000$\ell$) MnTe peaks indicates a highly oriented growth with the c-axis of $\alpha$-MnTe aligned perpendicular to the substrate surface, demonstrating epitaxial alignment of the film on GaAs(111). A high-resolution XRD scan over the $2\theta$ range of $53^\circ$--$58^\circ$, highlighting the MnTe(0004) and GaAs(222) reflections, is presented in Sec. 3 of the SM and Fig. S2(b).
Fig.~\ref{fig.fig1}(d) shows the evolution of RHEED patterns throughout the growth process. The pristine GaAs(111)B substrate exhibits sharp, streaky diffraction features, confirming a clean and well-ordered surface prior to deposition. Within 15 seconds after initiating growth, a slight shift in the streak positions accompanied by faint, diffuse features is observed, indicating the onset of film growth with lattice parameters distinct from those of the substrate. After 120 seconds of deposition, the initially streaky pattern breaks into spotty features, consistent with a three-dimensional growth mode at this stage. RHEED images acquired after 20 and 40 minutes reveal a gradual recovery of streak-like features from the spotty pattern, signifying the development of epitaxial ordering. After 55 minutes of growth, a clear streaky RHEED pattern is observed, indicating the formation of a smooth surface with long-range crystalline order. Additional RHEED patterns from several growth runs, including extended time-frame evolution during deposition, are provided in Sec. 2 of the SM and Fig. S1.

\begin{figure*} 
	\includegraphics[width=\linewidth]{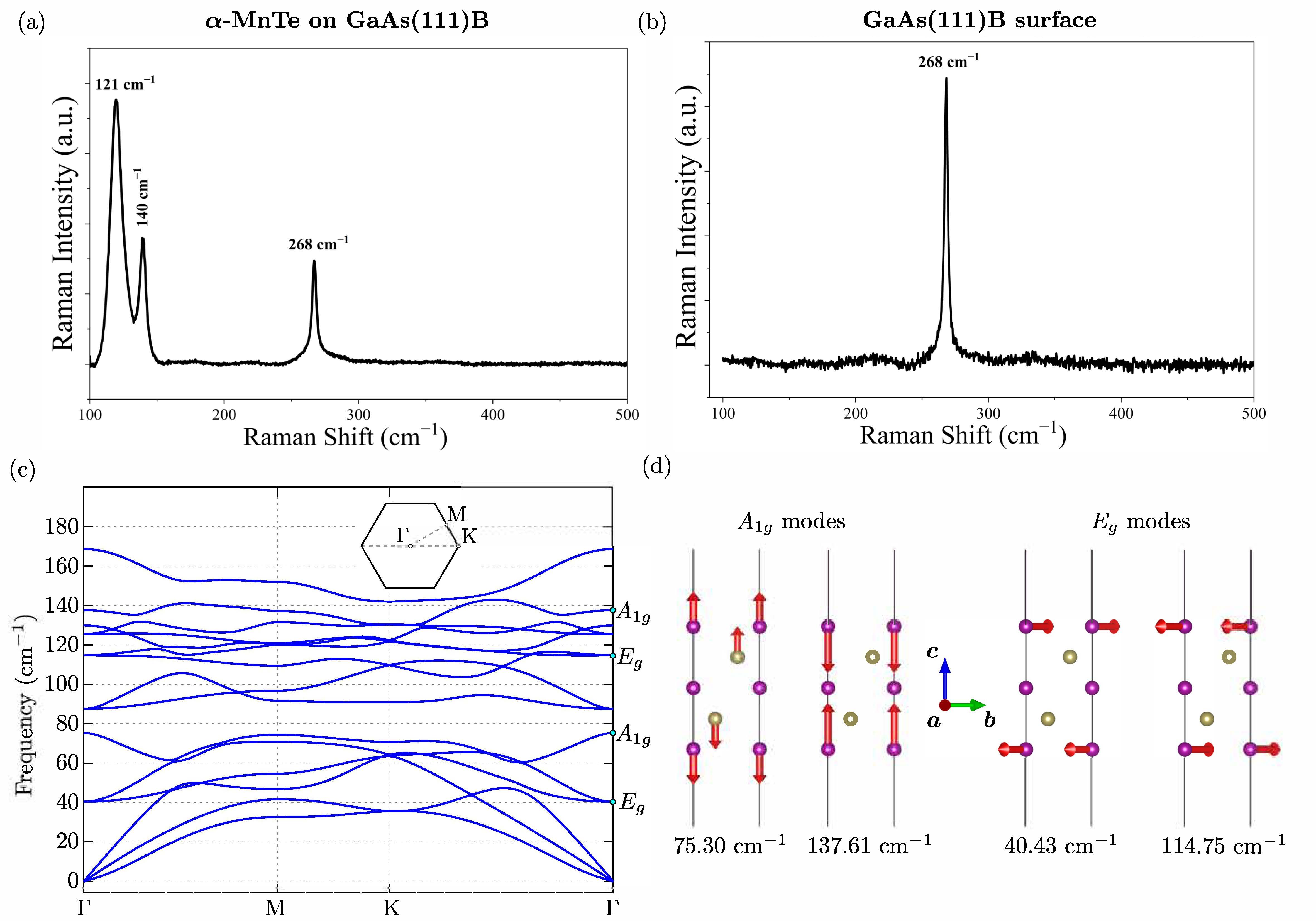} 
	\caption{
    {(a) Raman spectra of $\alpha$-MnTe film on GaAs(111)B surface. The results displayed three vibrational modes, from which $121$~cm$^{-1}$ and $140$~cm$^{-1}$ belong to the $\alpha$-MnTe and the $268$~cm$^{-1}$ peak belongs to the GaAs(111).
    (b) Raman spectra of bare GaAs(111)B surface.
    (c) Theoretically obtained phonon dispersion for $\alpha$-MnTe monolayer. 
    Inset show corresponding two dimensional Brillouin zone with high symmetry points.
    (d) Displacements given be the four Raman active modes for MnTe monolayer.
    Raman active modes are also marked by blue points on panel (c).}
\label{fig.fig3}}
\end{figure*}

Fig.~\ref{fig.fig2} presents the SEM and EDX results of the $\alpha$-MnTe/GaAs(111)B. Fig. ~\ref{fig.fig2}(a) shows the selected regions of the $\sim 10~\mu$m $\alpha$-MnTe thin film used for SEM imaging and EDX analysis.  The corresponding EDX elemental maps in Fig.~\ref{fig.fig2}(b) clearly identify Mn and Te from the deposited film along with strong Ga and As signals from the substrate. While Mn and Te signals are uniformly distributed across the surface the dominance of Ga and As indicates that the film is sufficiently thin for the substrate to contribute strongly to the EDX spectra. Quantitative composition data presented in Fig.~\ref{fig.fig2}(c) show Mn at 4.71 at.\% (3.50 wt.\%) and Te at 4.58 at.\% (7.90 wt.\%), while Ga and As contribute 47.19 at.\% (44.50 wt.\%) and 43.52 at.\% (44.10 wt.\%) respectively. Importantly, the data taken over several points, line scans, and selected regions all show consistent Mn and Te concentrations which confirms the uniform growth of $\alpha$-MnTe across the sample. Further SEM/EDS characterization, including elemental composition analysis, is presented in Sec. 3 of the SM and Fig. S2(a). These results verify the successful deposition of homogenous $\alpha$-MnTe films on GaAs(111)B. A near-ideal Mn:Te ratio ($\sim$1:1) from EDX matches the expected $\alpha$-MnTe stoichiometry, highlighting the effectiveness of flux calibration and substrate temperature during MBE growth. Notably, this near-ideal stoichiometry was achieved despite employing a Te-rich flux ratio ($\sim$1:6) during deposition, confirming that the excess Te compensates for its low sticking coefficient and surface desorption, eventually stabilizing stoichiometric $\alpha$-MnTe films. The compositional uniformity observed here is further supported by Raman spectroscopy and XRD measurements, which confirm the formation of the hexagonal $\alpha$-MnTe phase without detectable secondary phases.

{
Raman spectroscopy of $\alpha$-MnTe films grown on GaAs(111)B measured with 633 nm excitation, together with a reference spectrum from the bare substrate are presented in Fig.~\ref{fig.fig3}(a).
Obtained Raman spectra contain three distinct peaks, at $121$~cm${-1}$, $140$~cm${-1}$, and $268$~cm${-1}$.
First two peaks at $121~\mathrm{cm}^{-1}$ and $140~\mathrm{cm}^{-1}$ are assigned to the $E_{g}$ and $A_{1g}$ phonon modes of hexagonal $\alpha$-MnTe.
Such observation is in agreement with the previous Raman studies of MnTe~\cite{Li2022,zhang.raman}. 
The third prominent feature at $268~\mathrm{cm}^{-1}$ originates from the Raman active mode of the GaAs substrate.
Indeed, this can be confirmed by the reference spectrum of the bare GaAs(111)B surface, see Fig.~\ref{fig.fig3}(b).
Detailed experimental wavelength-dependent Raman spectroscopy measurements using 473 and 532 nm excitation wavelengths are presented in Sec. 4 of the SM and Figs. S3(a) and S3(b).}

The bulk $\alpha$-MnTe with P6$_{3}$/mmc symmetry contain only one Raman active modes $E_{g}$ (around $91.84$~cm$^{-1}$).
However, in the case of the $\alpha$-MnTe monolayer, the system symmetry is lowered to P$\bar{3}$m1 due to reduced dimensionality.
The obtained theoretically phonon dispersion curves for $\alpha$-MnTe monolayer is presented on Fig.~\ref{fig.fig3}(c).
In this case, there are four Raman active modes $2 A_{1g}$ ($75.30$~cm$^{-1}$ and $137.61$~cm$^{-1}$) and $2 E_{g}$ ($40.43$~cm$^{-1}$ and $114.75$~cm$^{-1}$), marked by green dots at $\Gamma$ point.
Thus, observed experimentally Raman peaks at $121$~cm$^{-1}$ and $140$~cm$^{-1}$ can be recognized as $E_{g}$ and $A_{1g}$ modes, respectively.
This modes arise from the collective vibrations of atoms within the hexagonal NiAs-type lattice, where the $E_{g}$ modes corresponds primarily to in-plane vibrations of Mn atoms, while the $A_{1g}$ modes involves out-of-plane oscillations along the c-axis [see Fig.~\ref{fig.fig3}(d)].

{
In our calculations we assume monolayer for which the primitive unit cell contain five atoms (three Mn and two Te). 
This structure lead to a stable two dimensional monolayer [see Fig.~\ref{fig.fig3}(c)]. 
In the case of another models (e.g. with four atoms), the layers were unstable in dynamical sense, due to imaginary phonon soft modes in the phonon spectra.
}

\begin{figure} 
	\includegraphics[width=\linewidth]{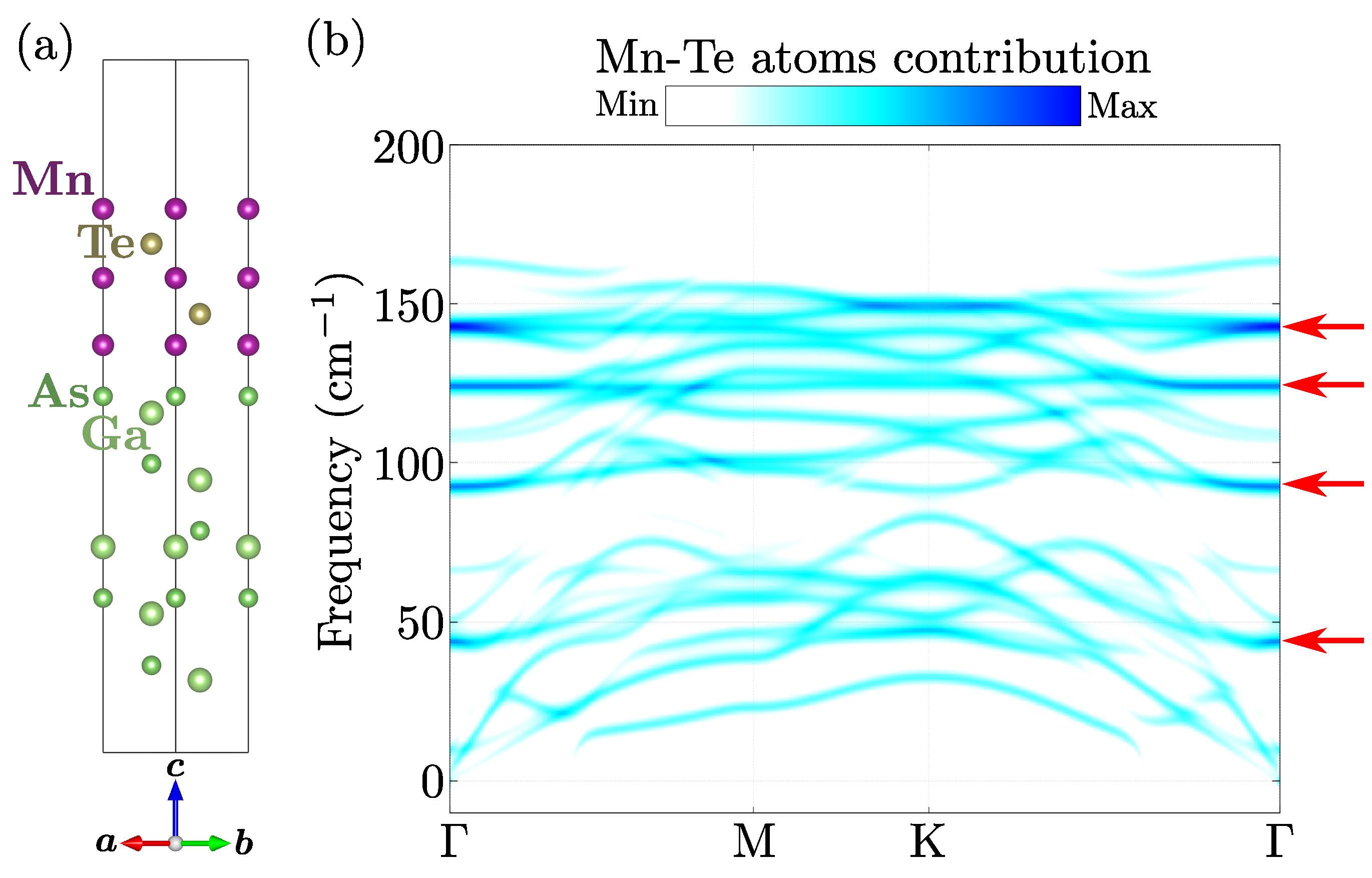} 
	\caption{{(a) Investigated $\alpha$-MnTe/GaAs(111)B interface. $\alpha$-MnTe monolayer is deposited on GaAs(111)B sufrace.
    (b) Projection of the phonon dispersion of the $\alpha$-MnTe/GaAs(111)B interface on the Mn-Te atoms contribution.}
\label{fig.fig4}}
\end{figure}

{
Situation can be more complicated in the case of the $\alpha$-MnTe/GaAs(111)B interface.
Here, due to the interplay between $\alpha$-MnTe monolayer and GaAs substrate, phonon modes described vibrations of all atoms.
Nevertheless, that GaAs exhibit one Raman active modes at $271.19$~cm$^{-1}$ (transverse optical) and $293.20$~cm$^{-1}$ (longitudinal optical)~\cite{strauch.gaas}.
Thus, the GaAs phonon modes at $\Gamma$ point are well separated from the MnTe vibrations.
To investigate this feature we construct $\alpha$-MnTe/GaAs(111)B interface modes, as presented on Fig.~\ref{fig.fig4}(a).
In this case, the primitive cell contain 15 atoms, which correspond to 45 phonon modes.
By analyses of the phonon polarization vectors, we can calculate the Mn-Te atoms contribution to the phonon sprectra [see Fig.~\ref{fig.fig4}(b)].
As we can see, at the $\Gamma$ point, the strong peaks corresponding to the $\alpha$-MnTe monolayer are well visible, e.g., around $44$~cm$^{-1}$, $90$~cm$^{-1}$, $125$~cm$^{-1}$ or $141$~cm$^{-1}$ (marked by red arrows).
Frequencies of such peaks very well satisfy observed Raman active modes.

In conclusion, we establish the successful epitaxial growth of high-quality $\alpha$-MnTe thin films on GaAs(111)B. Combined RHEED, XRD, and SEM/EDX measurements consistently confirm the high crystalline quality of the films through both \textit{in situ} and \textit{ex situ} characterization. Our Raman experiments, supported by first-principles phonon calculations, further reveal that the vibrational response of epitaxial $\alpha$-MnTe thin films is fundamentally distinct from that of bulk $\alpha$-MnTe. In particular, the reduced symmetry of the thin-film geometry and the $\alpha$-MnTe/GaAs(111)B interface affects the  Raman-visible phonons. The experimentally resolved modes near 121 and 140~cm$^{-1}$ establish the hexagonal NiAs-type phase and show that the interface-modified thin-film environment preserves distinct MnTe vibrational signatures while remaining well separated from the GaAs substrate response. These results identify symmetry lowering at the thin-film/interface level as a key factor in reshaping the phonon spectrum and provide direct access to the lattice dynamics of epitaxial $\alpha$-MnTe.}

\textbf{\textit{Acknowledgment}}: Work by M.N. at UCF was supported by the U.S. Department of Energy (DOE), Office of Science, Basic Energy Sciences (BES) under Award \# DE-SC0024304. Some figures in this work were rendered using {\sc Vesta}~\cite{momma.izumi.11} software.

\def\bibsection{\section*{\refname}}


\begin{thebibliography}{60}
\bibitem{Furdyna1988}
J. K. Furdyna,
Diluted magnetic semiconductors,
\href{https://doi.org/10.1063/1.366454}{\textit{J. Appl. Phys.} \textbf{64}, R29 (1988)}.

\bibitem{Zutic2004}
I. \v{Z}uti\'c, J. Fabian, and S. Das Sarma,
Spintronics: Fundamentals and applications,
\href{https://doi.org/10.1103/RevModPhys.76.323}{\textit{Rev. Mod. Phys.} \textbf{76}, 323 (2004)}.

\bibitem{Baltz2018}
V. Baltz, A. Manchon, M. Tsoi, T. Moriyama, T. Ono, and Y. Tserkovnyak,
Antiferromagnetic spintronics,
\href{https://doi.org/10.1103/RevModPhys.90.015005}{\textit{Rev. Mod. Phys.} \textbf{90}, 015005 (2018)}.

\bibitem{Hirohata2020}
A.~Hirohata, K.~Yamada, Y.~Nakatani, L.~Prejbeanu, B.~Diény, P.~Pirro, and B.~Hillebrands,
Review on spintronics: Principles and device applications,
\href{https://doi.org/10.1016/j.jmmm.2020.166711}
{\textit{J. Magn. Magn. Mater.} \textbf{509}, 166711 (2020)}.



\bibitem{Wolf2001}
S. A. Wolf, D. D. Awschalom, R. A. Buhrman, J. M. D. Coey, 
D. D. von Moln\'ar, M. L. Roukes, A. Y. Chtchelkanova, and D. M. Treger,
Spintronics: A spin-based electronics vision for the future,
\href{https://doi.org/10.1126/science.1065389}{\textit{Science} \textbf{294}, 1488 (2001)}.

\bibitem{Parkin2008}
S. S. P. Parkin, M. Hayashi, and L. Thomas,
Magnetic domain-wall racetrack memory,
\href{https://doi.org/10.1126/science.1145799}{\textit{Science} \textbf{320}, 190 (2008)}.


\bibitem{Nagaosa2010}
N. Nagaosa, J. Sinova, S. Onoda, A. H. MacDonald, and N. P. Ong,
Anomalous Hall effect,
\href{https://doi.org/10.1103/RevModPhys.82.1539}{\textit{Rev. Mod. Phys.} \textbf{82}, 1539 (2010)}.

\bibitem{Sinova2015}
J. Sinova, S. O. Valenzuela, J. Wunderlich, C. H. Back, and T. Jungwirth,
Spin Hall effects,
\href{https://doi.org/10.1103/RevModPhys.87.1213}{\textit{Rev. Mod. Phys.} \textbf{87}, 1213 (2015)}.









\bibitem{Jungwirth2016}
T. Jungwirth, X. Mart\'i, P. Wadley, and J. Wunderlich,
Antiferromagnetic spintronics,
\href{https://doi.org/10.1038/nnano.2016.18}{\textit{Nat. Nanotechnol.} \textbf{11}, 231 (2016)}.



\bibitem{Gomonay2014}
E.~V. Gomonay and V.~M. Loktev,
Spintronics of antiferromagnetic systems,
\href{https://doi.org/10.1063/1.4862467}
{\textit{Low Temp. Phys.} \textbf{40}, 17--35 (2014)}.


\bibitem{aa}
L.~\v{S}mejkal, J.~Sinova, and T.~Jungwirth,
Emerging research landscape of altermagnetism,
\href{https://doi.org/10.1103/PhysRevX.12.040501}
{\textit{Phys. Rev. X} \textbf{12}, 040501 (2022)}.

\bibitem{aaa}
L.~\v{S}mejkal, J.~Sinova, and T.~Jungwirth,
Beyond conventional ferromagnetism and antiferromagnetism:
A phase with nonrelativistic spin and crystal rotation symmetry,
\href{https://doi.org/10.1103/PhysRevX.12.031042}
{\textit{Phys. Rev. X} \textbf{12}, 031042 (2022)}.
\bibitem{aaaa}
L.~\v{S}mejkal, R.~Gonz\'alez-Hern\'andez, T.~Jungwirth, and J.~Sinova,
Crystal time-reversal symmetry breaking and spontaneous Hall effect
in collinear antiferromagnets,
\href{https://doi.org/10.1126/sciadv.aaz8809}
{\textit{Sci. Adv.} \textbf{6}, eaaz8809 (2020)}.

\bibitem{Krempasky2024}
J.~Krempaský, L.~Šmejkal, S.~W. D'Souza, M.~Hoffmann, J.~Bouaziz, B.~B. Sarma, D.~Kriegner, Y.~Hajlaoui, T.~W. Hänke, R.~González-Hernández, D.~Go, G.~Springholz, B.~Büchner, K.~Výborný, W.~Weber, G.~Woltersdorf, M.~Aeschlimann, J.~Minár, J.~Sinova, and T.~Jungwirth,
Altermagnetic lifting of Kramers spin degeneracy,
\href{https://doi.org/10.1038/s41586-023-06907-7}
{\textit{Nature} \textbf{626}, 517--522 (2024)}.

\bibitem{b}
I.~Mazin,
Editorial: Altermagnetism—A new punch line of fundamental magnetism,
\href{https://doi.org/10.1103/PhysRevX.12.040002}
{\textit{Phys. Rev. X} \textbf{12}, 040002 (2022)}.

\bibitem{bb}
S.-W.~Cheong and F.-T.~Huang,
Altermagnetism with non-collinear spins,
\href{https://doi.org/10.1038/s41535-024-00626-6}{\textit{npj Quantum Mater.} \textbf{9}, 13 (2024)}.



\bibitem{bbb}
R.~D.~Gonzalez Betancourt \textit{et al.},
Spontaneous anomalous Hall effect arising from an unconventional
compensated magnetic phase in a semiconductor,
\href{https://doi.org/10.1103/PhysRevLett.130.036702}
{\textit{Phys. Rev. Lett.} \textbf{130}, 036702 (2023)}.
\bibitem{Guo2023}
Y.~Guo, H.~Liu, O.~Janson, I.~C.~Fulga, J.~van~den~Brink, and J.~I.~Facio,
Spin-split collinear antiferromagnets: A large-scale ab-initio study,
\href{https://doi.org/10.1016/j.mtphys.2023.100991}
{\textit{Mater.\ Today Phys.} \textbf{32}, 100991 (2023)}.

\bibitem{Mazin2023}
I.~Mazin, R.~Gonz\'alez-Hern\'andez, and L.~\v{S}mejkal,
Induced monolayer altermagnetism in MnP(S, Se)$_3$ and FeSe,
\href{https://arxiv.org/abs/2309.02355}
{arXiv:2309.02355 (2023)}.

\bibitem{Regmi2024}
R.~B.~Regmi \textit{et al.},
Altermagnetism in the layered intercalated transition metal dichalcogenide CoNb$_4$Se$_8$,
\href{https://doi.org/10.1038/s41467-025-58642-4}
{\textit{Nat. Commun.} \textbf{16}, 4399 (2025)}.

\bibitem{Dong2025}
J.~Dong, K.~Wu, M.~Zhu, F.~Zheng, X.~Li, and J.~Zhang,
Nonrelativistic spin-splitting multiferroic antiferromagnet and compensated ferrimagnet with zero net magnetization,
\href{https://arxiv.org/abs/2501.02914}
{arXiv:2501.02914 (2025)}.
\bibitem{Sakhya2025}
A.~P.~Sakhya, M.~I.~Mondal, M.~Sprague, R.~B.~Regmi, A.~K.~Kumay, H.~Sheokand, I.~I.~Mazin, N.~J.~Ghimire, and M.~Neupane,
Electronic structure of a layered altermagnetic compound CoNb$_4$Se$_8$,
\href{https://arxiv.org/abs/2503.16670}
{arXiv:2503.16670 (2025)}.

\bibitem{Sprague2025}
M.~Sprague, M.~I.~Mondal, A.~P.~Sakhya, R.~B.~Regmi, S.~Sadhukhan, A.~K.~Kumay, H.~Sheokand, I.~I.~Mazin, N.~J.~Ghimire, and M.~Neupane,
Observation of altermagnetic spin-splitting in an intercalated transition metal dichalcogenide,
\href{https://arxiv.org/abs/2508.12985}
{arXiv:2508.12985 (2025)}.

\bibitem{Janik1995}
E.~Janik, E.~Dynowska, J.~Bąk-Misiuk, M.~Leszczyński, W.~Szuszkiewicz, T.~Wojtowicz, G.~Karczewski, A.~K. Zakrzewski, and J.~Kossut,
Structural properties of cubic MnTe layers grown by MBE,
\href{https://doi.org/10.1016/0040-6090(95)06632-2}
{\textit{Thin Solid Films} \textbf{267}, 74--78 (1995)}.

\bibitem{ElectricalProperties2016}
M. E. Vickers, D. A. MacLaren, and G. A. Gehring,
Electrical properties of NiAs-type MnTe films with preferred orientation,
\href{https://doi.org/10.1063/1.4942833}{\textit{J. Appl. Phys.} \textbf{119}, 045304 (2016)}.
\bibitem{Osumi2024}
T. Osumi, S. Souma, T. Aoyama, K. Yamauchi, A. Honma, K. Nakayama, T. Takahashi, K. Ohgushi, and T. Sato,
Observation of giant band splitting in altermagnetic MnTe,
\href{https://doi.org/10.1103/PhysRevB.109.115102}{\textit{Phys. Rev. B} \textbf{109}, 115102 (2024)}.

\bibitem{Roessler2025}
S. Rößler, V. Ginga, M. Schmidt, Y. Prots, H. Rosner, U. Burkhardt, U. K. Rößler, and A. A. Tsirlin,
Low-field magnetization processes of hexagonal easy-plane altermagnet $\alpha$-MnTe,
\href{https://arxiv.org/abs/2511.01388}{arXiv:2511.01388 (2025)}.
\bibitem{Gonzalez2024}
R. D. González Betancourt, J. Zubáč, K. Geishendorf, P. Ritzinger, B. Růžičková, T. Kotte, J. Železný, G. Springholz, B. Büchner, A. Thomas, H. Reichlová, D. Kriegner, and T. Jungwirth,
Anisotropic magnetoresistance in altermagnetic MnTe,
\href{https://doi.org/10.1038/s44306-024-00046-z}{\textit{Nat. Rev. Phys.} \textbf{6}, 683 (2024)}.

\bibitem{Osumi2024band}
T. Osumi \textit{et al.},
Observation of giant band splitting in altermagnetic MnTe,
\href{https://doi.org/10.1103/PhysRevB.109.115102}{\textit{Phys. Rev. B} \textbf{109}, 115102 (2024)}.



\bibitem{d}
K.~P.~Kluczyk \textit{et al.},
Coexistence of anomalous Hall effect and weak magnetization
in a nominally collinear antiferromagnet MnTe,
\href{https://doi.org/10.1103/PhysRevB.110.155201}
{\textit{Phys. Rev. B} \textbf{110}, 155201 (2024)}.
\bibitem{c}
R.~Watanabe, R.~Yoshimi, M.~Shirai, T.~Tanigaki, M.~Kawamura,
A.~Tsukazaki, K.~S.~Takahashi, R.~Arita, M.~Kawasaki, and Y.~Tokura,
Emergence of interfacial conduction and ferromagnetism in MnTe/InP,
\href{https://doi.org/10.1063/1.5052635}
{\textit{Appl. Phys. Lett.} \textbf{113}, 181602 (2018)}.

\bibitem{cc}
M.~Chilcote \textit{et al.},
Stoichiometry-induced ferromagnetism in altermagnetic candidate MnTe,
\href{https://doi.org/10.1002/adfm.202405829}
{\textit{Adv. Funct. Mater.} \textbf{34}, 2405829 (2024)}.

\bibitem{e}
J.~D. Wasscher,
Evidence of weak ferromagnetism in MnTe from galvanomagnetic measurements,
\href{https://doi.org/10.1016/0038-1098(65)90284-X}
{\textit{Solid State Commun.} \textbf{3}, 169--171 (1965)}.


\bibitem{eee}
S.~Bey \textit{et al.},
Unexpected tuning of the anomalous Hall effect in altermagnetic MnTe thin films,
\href{https://arxiv.org/abs/2409.04567}
{arXiv:2409.04567 (2024)}.
\bibitem{Bey2025}
S.~Bey, M.~Zhukovskyi, T.~Orlova, S.~S.~Fields, V.~Lauter, H.~Ambaye, A.~Ievlev, S.~P.~Bennett, X.~Liu, and B.~A.~Assaf,
Interface, bulk and surface structure of heteroepitaxial altermagnetic $\alpha$-MnTe films grown on GaAs(111),
\href{https://doi.org/10.1103/1dh1-mly1}
{\textit{Phys. Rev. Mater.} \textbf{9}, 074404 (2025)}.
\bibitem{Chen2026}
A.H.~Chen, P.~R. Raghuvanshi, J.~Cook, M.~Chilcote, J.~Lapano, A.~R. Mazza, Q.~Lu, S.~Kim, Y.-C.~Wu, T.~Z. Ward, B.~J. Lawrie, G.~Bian, J.~Burns, J.~D. Poplawsky, M.-G.~Han, Y.~Zhu, L.~Lindsay, H.~Miao, Z.~Gai, R.~G. Moore, G.~Eres, V.~R. Cooper, and M.~Brahlek,
Programmable phase selection between altermagnetic and noncentrosymmetric polymorphs of MnTe on InP via molecular beam epitaxy,
\href{https://doi.org/10.1021/acsami.5c14582}
{\textit{ACS Appl. Mater. Interfaces} \textbf{18}, 15654--15664 (2026)}.

\bibitem{Bangar2025}
H.~Bangar, P.~Tsipas, P.~Rout, L.~Pandey, A.~Kalaboukhov, A.~Lintzeris, A.~Dimoulas, and S.~P. Dash,
Interplay between altermagnetic order and crystal symmetry probed using magnetotransport in epitaxial altermagnet MnTe,
\href{https://doi.org/10.48550/arXiv.2505.14589}
{arXiv:2505.14589 (2025)}.

\bibitem{Li2022}
S.~Li, J.~Wu, B.~Liang, L.~Liu, W.~Zhang, N.~Wazir, J.~Zhou, Y.~Liu, Y.~Nie, Y.~Hao, P.~Wang, L.~Wang, Y.~Shi, and S.~Li,
Antiferromagnetic $\alpha$-MnTe: Molten-Salt-Assisted Chemical Vapor Deposition Growth and Magneto-Transport Properties,
\href{https://doi.org/10.1021/acs.chemmater.1c04066}
{\textit{Chem. Mater.} \textbf{34}, 873 (2022)}.

\bibitem{zhang.raman}
J. Zhang, Q. Lian, Z. Pan, W. Bai, J. Yang, Y. Zhang, X. Tang, and J. Chu,
Spin-phonon coupling and two-magnons scattering behaviors in hexagonal NiAs-type antiferromagnetic MnTe epitaxial films,
\href{https://doi.org/10.1002/jrs.5928}{{\it J Raman Spectrosc.} {\bf 51}, 1383 (2020)}.

\bibitem{strauch.gaas} D. Strauch and B. Dorner,
Phonon dispersion in GaAs,
\href{http://doi.org/10.1088/0953-8984/2/6/006}{{\it J. Phys.: Condens. Matter} {\bf 2}, 1457 (1990)}.

\bibitem{Kommichau1986}
G.~Kommichau, H.~Neumann, and W.~Schmitz,
Thermal expansion of SrF$_2$ at elevated temperatures,
\href{https://doi.org/10.1002/crat.2170211221}
{\textit{Cryst.\ Res.\ Technol.} \textbf{21}, 1583 (1986)}.

\bibitem{blochl.94}  P. E. Bl\"{o}chl, Projector augmented-wave method, \href{https://doi.org/10.1103/PhysRevB.50.17953}{{\it Phys. Rev. B} {\bf 50}, 17953 (1994)}.

\bibitem{kresse.hafner.94} G. Kresse and J. Hafner, Ab initio molecular-dynamics
simulation of the liquid-metal–amorphous-semiconductor transition in germanium, \href{https://doi.org/10.1103/PhysRevB.49.14251}{{\it Phys. Rev. B} {\bf 49}, 14251 (1994)}.

\bibitem{kresse.furthmuller.96} G. Kresse and J. Furthm\'{u}ller, Efficient iterative schemes for ab initio total-energy calculations using a plane-wave basis set, \href{https://doi.org/10.1103/PhysRevB.54.11169}{{\it Phys. Rev. B} {\bf 54}, 11169 (1996)}.

\bibitem{kresse.joubert.99}  G. Kresse and D. Joubert, From ultrasoft pseudopotentials to the projector augmented-wave method, \href{https://doi.org/10.1103/PhysRevB.59.1758}{{\it Phys. Rev. B} {\bf 59}, 1758 (1999)}.

\bibitem{perdew.burke.96} J. P. Perdew, K. Burke, and M. Ernzerhof, Generalized
gradient approximation made simple, \href{https://doi.org/10.1103/PhysRevLett.77.3865}{{\it Phys. Rev. Lett.} {\bf 77}, 3865 (1996)}.

\bibitem{liechtenstein.anisimov.95} A. I. Liechtenstein, V. I. Anisimov, and J. Zaanen, Density-functional theory and strong interactions: Orbital ordering in Mott--Hubbard insulators, \href{https://doi.org/10.1103/PhysRevB.52.R5467}{{\it Phys. Rev. B} {\bf 52}, R5467 (1995)}.

\bibitem{monkhorst.pack.76} H. J. Monkhorst and J. D. Pack, Special points for Brillouin-zone integrations, \href{https://doi.org/10.1103/PhysRevB.13.5188}{{\it Phys. Rev. B} {\bf 13}, 5188 (1976)}.

\bibitem{togo.chaput.23} A. Togo, L. Chaput, T. Tadano, and I. Tanaka, Implementation strategies in phonopy and phono3py, \href{http::/doi.org/10.1088/1361-648X/acd831}{{\it J. Phys. Condens. Matter} {\bf 35}, 353001 (2023)}.

\bibitem{togo.23} A. Togo, First-principles phonon calculations with phonopy and phono3py, \href{https://doi.org/10.7566/JPSJ.92.012001}{{\it J. Phys. Soc. Jpn.} {\bf 92}, 012001 (2023)}.

\bibitem{momma.izumi.11} K. Momma and F. Izumi, vesta3 for three-dimensional
visualization of crystal, volumetric and morphology data, \href{https://doi.org/10.1107/S0021889811038970}{{\it J. Appl. Crystallogr.} {\bf 44}, 1272 (2011)}.

\end{thebibliography}
\end{document}